\documentclass[aps,preprint]{revtex4}%
\usepackage{amsfonts}
\usepackage{amsmath}
\usepackage{amssymb}
\usepackage{graphicx}%
\begin{document}
\preprint{Internal Communication }
\title{Evidence for a dark matter particle}
\author{Yukio Tomozawa}
\affiliation{Michigan Center for Theoretical Physics}
\affiliation{Randall Laboratory of Physics, University of Michigan}
\affiliation{Ann Arbor, MI. 48109-1040, USA}
\date{\today }

\begin{abstract}
A prediction and observational evidence for the mass of a dark matter particle
are presented..

\end{abstract}

\pacs{04.70.-s, 95.85.Pw, 95.85.Ry, 98.54.Cm}
\maketitle

\section{\label{sec:level1}Introduction}

It has been reported by HESS that there are excess gamma rays above the TeV
range from some gamma ray emitters. In the PKS 2005-489 gamma ray data, excess
gamma rays appear above 2 TeV, which is above the power law extension of the
low energy data\cite{hess}. However, it is difficult to draw a conclusion from
data for an individual object due to poor statistics. If one is interested in
gamma rays from dark matter particle (DMP) and antiparticle annihilation, one
can combine data from various sources, since a sharp peak at the same energy
can be expected. The only necessary procedure is statistical weighing to sum
up data of different quality. To prove a point, I have selected part of the
HESS data obtained from 8 sources which are located in neighboring space and
time and have comparable statistics, with an energy range of 1 to 20 TeV.

\section{Theoretical prediction for DMP mass}

Recent data from the Pierre Auger Project has reported a strong correlation
between \ sources of high energy cosmic rays and AGN\cite{auger}. The author
has predicted this data since 1985. In his model, the author has shown that
quantum field theory yields a repulsive component at short distances for
gravity and that black hole collapse results in an explosion by repulsive
forces. The knee energy of the cosmic ray energy spectrum is an inter-phase of
radiation-dominated expansion and matter-dominated expansion, similar to the
expansion of the universe\cite{cr1},\cite{cr2},\cite{cr3}. This necessitates
the existence of a mass scale at 3 PeV, the knee energy of cosmic rays. This
mass scale may be an indication of new physics. If it corresponds to a
supersymmetric multiplet, the lowest mass particle (LMP) is a candidate for a DMP.

In order to have a mass scale of 3 PeV and a DMP of relatively low mass, one
has to have a supersymmetry with a large mass ratio. Such a theory has been
proposed by GLMR (Giudice, Luty, Murayama and Rattazzi). Assuming the absence
of singlets, GLMR derived a large mass ratio\cite{glmr},%
\begin{equation}
M_{2}=\frac{\alpha}{4\pi\sin^{2}\theta_{W}}m_{3/2}=2.7\times10^{-3}m_{3/2},
\end{equation}
among other parameter relations, where $\alpha$ and $\theta_{W}$ are the fine
structure constant and the weak interaction angle respectively. Since this is
the largest mass ratio obtained, one may choose the highest mass scale to be%
\begin{equation}
m_{3/2}=3\text{ }PeV;
\end{equation}
then one gets\cite{cr4}%
\begin{equation}
M_{2}=8.1\text{ }TeV,
\end{equation}
which is the mass of LMP, i.e., the DMP mass. Being a weakly interacting
particle, this particle must be produced by a pair in an accelerator
experiment. This makes it impossible to discover these particles directly in
LHC experiments at the presently planned energy scale. The accuracy of the
prediction is in the range of 10 \% from the determination of the knee energy
of the cosmic ray energy spectrum.

\section{TeV gamma ray data}

In a recent HESS report\cite{hess1}, high energy gamma rays from 8 unknown
sources have been recorded. The data from each source cover the energy range
of 1 to 20 TeV and have similar statistics, since they have been obtained in a
recent systematic survey. That the sources are unknown may not be a drawback
for a dark matter gamma ray search, since unknown sources may not be ordinary
AGN or other known astronomical objects. If the source is an AGN type object
consisting entirely of dark matter particles, it may not have the signature of
an ordinary AGN, since such a signature needs ordinary matter to emit atomic
photons. The presence of abundant dark matter favors 2 gamma ray emission from
DMP and anti-DMP annihilation. One does not need to exclude gamma ray emitters
such as ordinary AGN etc, since one expects a DMP environment in such a case
as well. The simple sum of gamma rays from the 8 sources is plotted in Fig. 1.
The error bars are estimated from the existing data. The sum clearly showes a
peak at 7.6 $\pm$ 0.1 TeV. This is consistent with the predicted value of 8.1
$\pm$ 0.8 TeV.

Clearly, I have chosen a small subset of HESS data. In order to get a more
accurate estimate for this peaking phenomenon, it is desirable to analyze much
more data. Here are some suggestions.

1) In order to get a sensible conclusion, one has to sum up data from many
sources, as is done in this article. For a DMP + anti-DMP $\rightarrow2\gamma$
search, this is a reasonable approach.

2) Choose all data with an exceess in the 1-15 TeV energy range. Data that do
not have excess in multi-TeV range contribute only to increased statistical
error bars without adding a significant contribution to the peak value..

3) Sum up data in the 1-15 TeV range, since the error beyond this energy range
tends to increase due to the scarcity of events.

4) The sources can be identifid or unidentified. The latter should not be
excluded, since there is a chance to have more dark matter contributions from
such events.

5) A systematic survey is welcome, since such an approach tends to have a
consistent statistical significance.

It is hoped that this report will encourage a systematic DMP mass search in
the 1-15 TeV energy range.

\begin{acknowledgments}
The author would like to thank David N. Williams for useful discussion and
reading the manuscript.\bigskip
\end{acknowledgments}

\begin{quote}
Correspondence should be addressed to the author at tomozawa@umich.edu.
\end{quote}

Figure caption

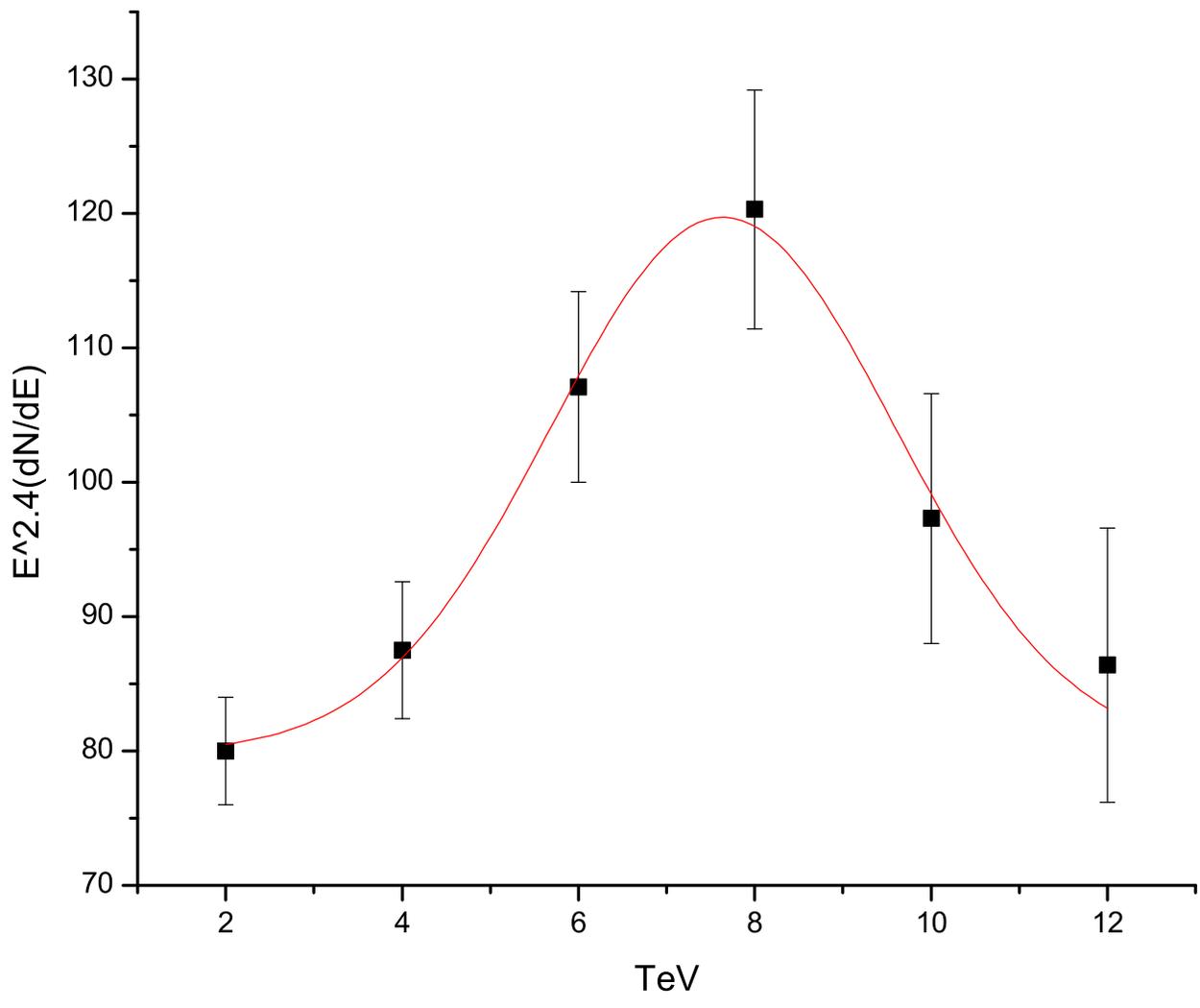
\bigskip Fig. 1. Sum of gamma ray energy spectra of 8 unidentified
sources\cite{hess1}. The y axis is E$^{2.4}(dN/dE)$ in units of 10$^{-12}%
$(TeV)$^{0.4}$(erg cm$^{-2}$s$^{-1}$).

\end{document}